\documentclass[epj,twocolumn]{webofc}
\usepackage[varg]{txfonts}   
\woctitle{Powders \& Grains 2017}
\begin{document}
\title{Percolation study for the capillary ascent of a liquid through a granular soil}
%
%

\author{\firstname{M. A.} \lastname{C\'ardenas-Barrantes}\inst{1}\fnsep\thanks{\email{macardenasb@unal.edu.co}} \and
        \firstname{J. D.} \lastname{ Mu\~{n}oz}\inst{1}\fnsep\thanks{\email{jdmunozc@unal.edu.co}} \and
        \firstname{N. A. M.} \lastname{Araujo}\inst{2}\fnsep\thanks{\email{nmaraujo@fc.ul.pt}}
}

\institute{Simulation of Physical Systems Group, Department of Physics, Universidad Nacional de Colombia, Carrera 30 No. 45-03, Ed. 404, Of. 348, Bogota D.C., Colombia.
\and
           Centro de Física Teórica e Computacional, Departamento de Física, Faculdade de Ciências, Universidade de Lisboa, Campo Grande, P-1749-016 Lisboa, Portugal.
          }

\abstract{%
Capillary rise plays a crucial role in the construction of road embankments in flood zones, where hydrophobic compounds are added to the soil to suppress the rising of water and avoid possible damage of the pavement. Water rises through liquid bridges,  menisci and trimers, whose width and connectivity depends on the maximal half-length $\lambda$ of the capillary bridges among grains. Low $\lambda$s generate a disconnect structure, with small clusters everywhere. On the contrary, for high $\lambda$, create a percolating cluster of trimers and enclosed volumes that form a natural path for capillary rise. Hereby, we study the percolation transition of this geometric structure as a function of $\lambda$ on a granular media of monodisperse spheres in a random close packing. We determine both the percolating threshold $\lambda_c=(0.049\pm 0.004)R$ (with R the radius of the granular spheres), and the critical exponent of the correlation length $\nu=0.830\pm 0.051$, suggesting that the percolation transition falls into the universality class of ordinary percolation.
}
\maketitle
\section{Introduction}
\label{Introduction}
Consider a sandpile with its base covered with water. If grains are close enough, water will rise through the interstices of the grains. This capillary rise plays a major role in the transport of fluids across porous media, including water and oil wells. Specially in unsaturated soil structures, like embankments, the capillary rise of water is a real concern, because water can damage the integrity of the structure \cite{_Unsa2014}. The broad spectrum of possible solutions include the use of hydrophobic materials \cite{_Unsa2014} or even the addition of active mechanisms to compensate for the deformations produced by capillary forces \cite{Pozzato_Char2014}. 

The rising of water through a granular medium is strongly determined by the geometry of the interconnected structure of pores among the grains. First models \cite{Fatt_The_1956} represented that structure by  sites (pore bodies) of arbitrary shape and position interconnected by bonds (pore throats), whose sizes and shapes could be obtained from experimental probes \cite{Vogel_Quant_2001,Coles_Pore1998}; many properties, like relative permeability \cite{Fischer_Pred_1999}  or drainage and imbibition \cite{Mason_Simu1995} can be estimated from this simplified model. More recent works focus on 
representing the liquid structures among grains (bridges, menisci and pore bodies) as real as possible, based on experiments (as in X-Ray microtomography \cite{Scheel_Morp2008}) or in computer simulations \cite{Motealleh_Unif2013, Melnikov_Grai2015}. This modeling is able to reproduce water saturation and drying with the water volume as control parameter and to compute forces and pressures \cite{Melnikov_Grai2015}. Together with experiments, they have succeeded identifying trimers (that is, the junction of three liquid bridges and a meniscus) as the minimal building block to build a pathway for rising water \cite{Melnikov_Grai2015, Scheel_Morp2008}. The structure itself must strongly change with the critical length of a capillary bridge (a function of contact angle and surface tension \cite{Megias_Capil2009}), and some studies in two dimensions have been performed to find when a connected structure first appears as either the contact angle \cite{Martys_Crit1991} or the liquid volume \cite{Cieplak_Dyna1988, Berkowitz_Perc1993} increases.

The present work investigates how the set of trimers and enclosing pore bodies at disposal for capillary rising changes from a fully disconnected structure to a connected pathway as the critical length of the capillary bridges increases. The goal is to find both the critical half-length $\lambda_c$ for the transition and the critical exponent $\nu$ for the correlation length. The study is performed on the interstices on three-dimensional random close packings of monodisperse spheres. The capillary model of trimers and pore bodies on random close packings is introduced in Sec. \ref{SecCapillarymodel}. Next, Sec. \ref{SecProcedure} analyses the resulting capillary structures by using tools of percolation theory \cite{Stauffer1994,Sahimi_Appl1994,Araujo_Rece2014}. Finally, Sec. \ref{SecConclusions} summarizes the main conclusions and discussions.      

\begin{figure}[h]
\centering
  \includegraphics[height=6cm]{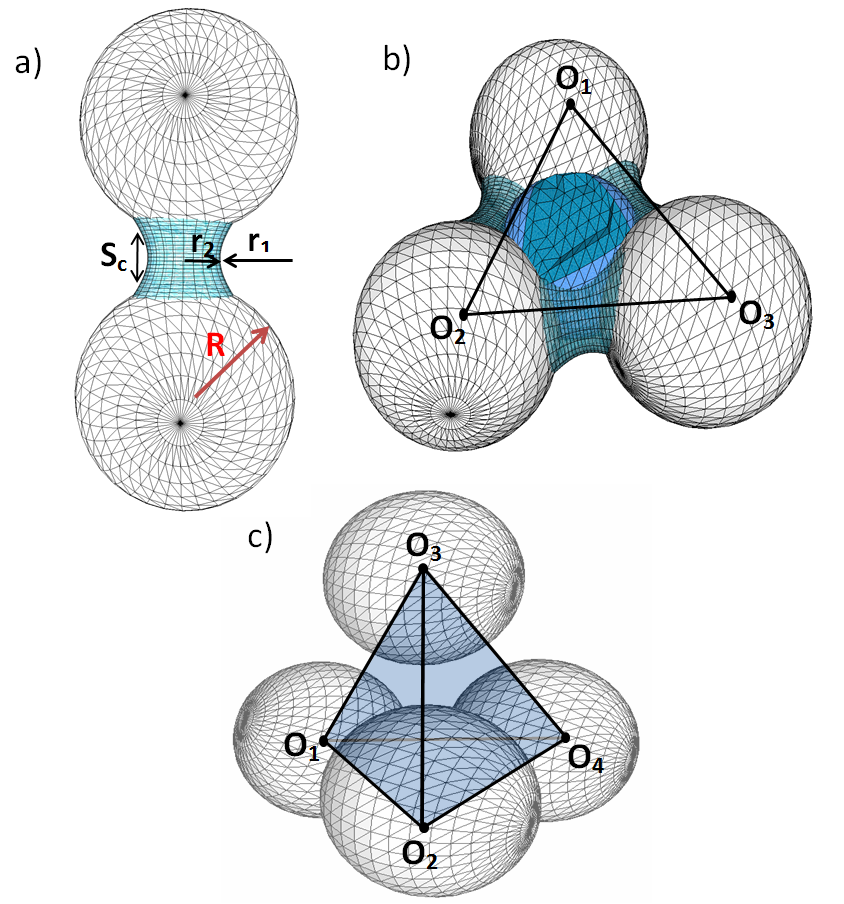}
\caption{Main capillary structures. (a) Liquid bridge with separation distance $S$. (b) Trimer, built by three liquid bridges (light blue) and a meniscus (dark blue). (c) A liquid volume enclosed by four trimers.}
  \label{fgr:fig2}
\end{figure} 

\section{Capillary model} 
\label{SecCapillarymodel}

\begin{figure}[h]
\centering
  \includegraphics[height=4.5cm]{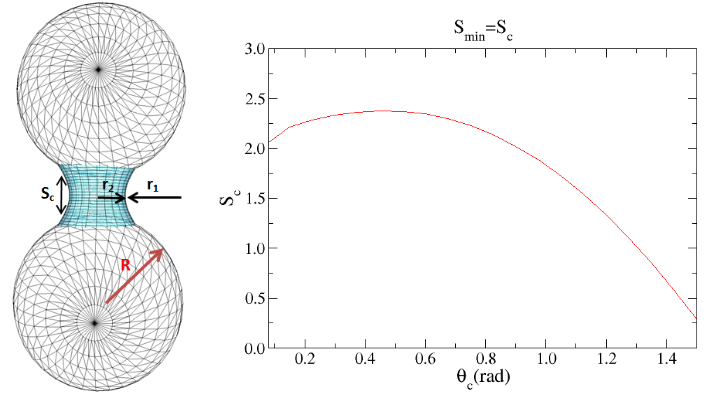}
\caption{(a) Liquid bridge with separation distance $S$. (b) Maximal separation distance $S_c$ for a capillary bridge as function of the contact angle $\theta$ in the toroidal approximation (deduced from data in \cite{Megias_Capil2009}).}
  \label{fgr:TorAp}
\end{figure}

From a microscopic point of view, water can rise if the grains are close enough to build capillary bridges among them. The shape of a capillary bridge between two identical spherical grains is not strongly affected by gravity, but determined by the contact angle $\theta_c$, the liquid-gas surface tension $\gamma$, the liquid volume $V$ and the distance $S$ between the grains \cite{Guopingcap4}. There is a maximal distance among grains $S_c=2\lambda$ a capillary bridge can overcome. If $S<2\lambda$, a capillary bridge may eventually established for some $V$; otherwise, there is no possible path between the grains for the water to rise. This limit can be estimated through the toroidal approximation method \cite{Fisher1926} (Fig. \ref{fgr:TorAp}) or by numerical simulations.

The bridges among grains should be connected in order to build a path for the rising water. From a theoretical point of view, it would be possible to join two bridges with a meniscus; but micro-tomographies on experimental random granular arrays of monodisperse spheres \cite{Scheel_Morp2008} do not show that kind of structure. On the contrary, the minimal connected structures are trimers, that are the junction of three liquid bridges and a meniscus \cite{Melnikov_Grai2015} (Fig. \ref{fgr:fig2}). A trimer will eventually form for some water content if three grains are so close together that the distances between any two of them are shorter than $S_c$ and the angles between every two bridges are smaller than $\pi - 2\theta_c$ \cite{Megias_Capil2009}. Two trimers are assumed connected if they share a bridge, and connected trimers can eventually enclose filling volumes (Fig. \ref{fgr:fig2}). Trimers and enclosed volumes form the structure for capillary rising. If $S_c$ is small, the structure is a disconnected set of clusters. On the contrary, if $S_c$ is large enough, there is a connected path across the sample, i.e.  a percolating passage for the liquid to rise. The aim of the present work is to characterize the  transition across these two regimes \-- driven by $\lambda$, the half maximal length of a capillary bridge \-- on a monodisperse set of spherical grains by using the standard tools of classical percolation.

\begin{figure}[h]
\centering
  \includegraphics[height=4.5cm]{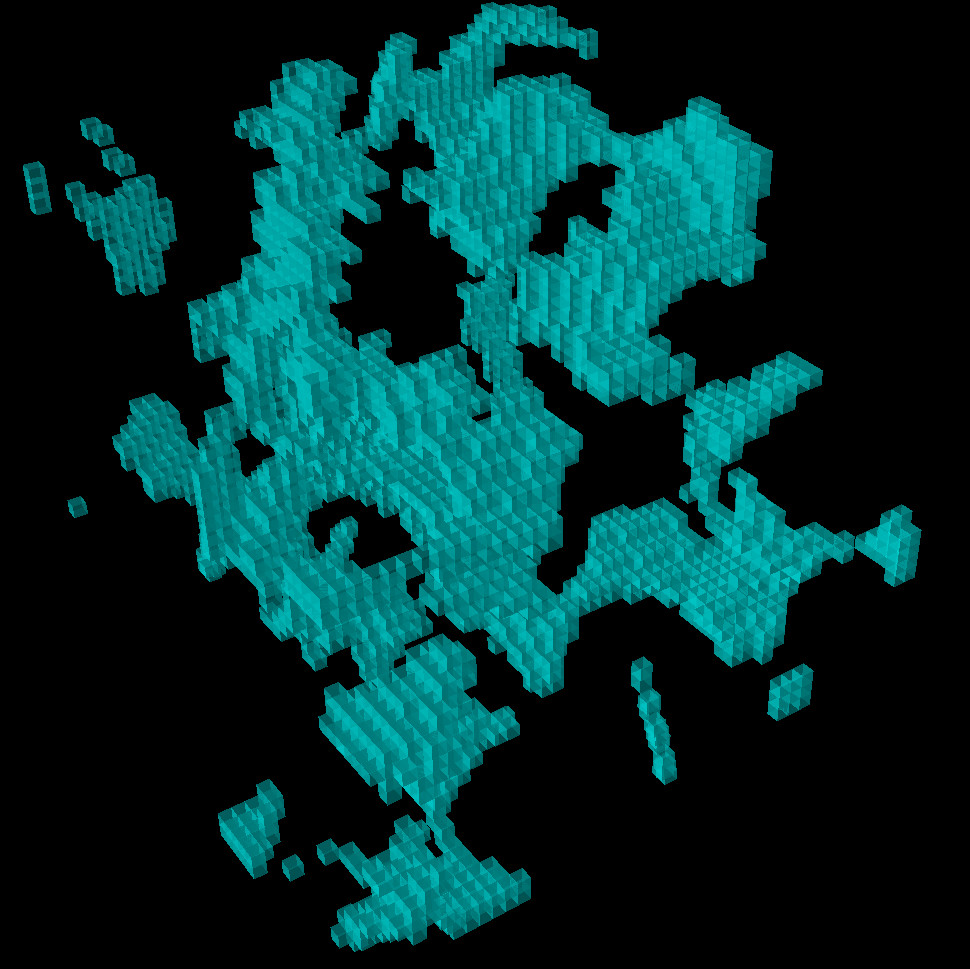}
\caption{Three-dimensional plot of all capillary structures (bridges, trimers and enclosed volumes) in a mono-disperse sample of 87 grains with volume fraction $\phi\simeq0.63$ at a capillary length $\lambda=0.025R$, $R$ the radius of the grains.}
  \label{fgr:TetraCell}
\end{figure}

\section{Procedure and Results} 
\label{SecProcedure}

The granular medium was modeled as a dense random packing of monodisperse spheres of unitary radius with volume fraction $\phi \approx 0.63$, slightly below the random close packing \cite{Shahinpoor_An1980}. The spatial configurations of spheres were generated by using the algorithm of Jodrey and Tory \cite{Jodrey_Comp1985} on cubic volumes of sizes $L=22,26,30,36,42,48,56$ and $64$, and configurations were accepted if the overlapping lengths between any two spheres was below $0.015$, in units of particle radius. With this procedure, 200 configurations per size for sizes $L\le 42$ and 500 configurations per size for sizes $L\ge 48$ were generated, corresponding to around 1560 grains each for $L=22$ and 38200 grains each for $L=64$. 

\begin{figure}[h]
\centering
  \includegraphics[height=6.5cm]{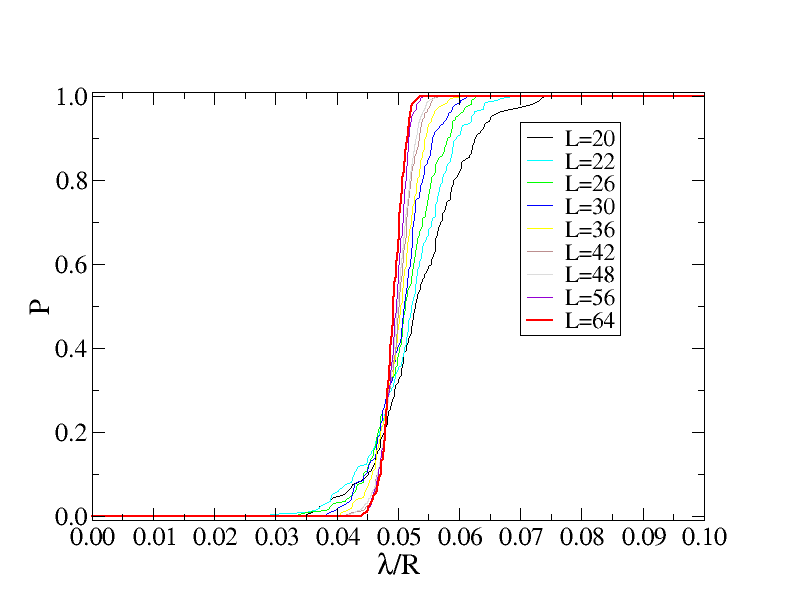}
\caption{Probability of the capillary structure to percolate as a function of the maximal half-length of a capillary bridge $\lambda$.}
  \label{fgr:P}
\end{figure}

For each configuration, the system was discretized with a fine grid of cubic cells of side $a=0.031$. Once a value for $\lambda$ was set, each cell was checked to be part of a trimer or an enclosed volume. Two cells are said to be part of the same cluster if they are first neighbors. Next, the Directional Propagation Labelling algorithm (DPL) \cite{Hawick_Para2010}, implemented on GPUs, was employed to determine if there was a percolating cluster connecting the lower and upper borders of the cube. Finally, a bisection algorithm \cite{Press_Nume2007} was implemented to determine the effective critical value $\lambda_c^{\rm eff}$ where a percolating cluster first appears on that configuration. The cumulative distribution of these effective critical values for each size are the sigmoids (Fig. \ref{fgr:P}) that will be the incoming data for the finite-size scaling procedure that follows. 

\begin{figure}[h]
\centering
  \includegraphics[height=6.5cm]{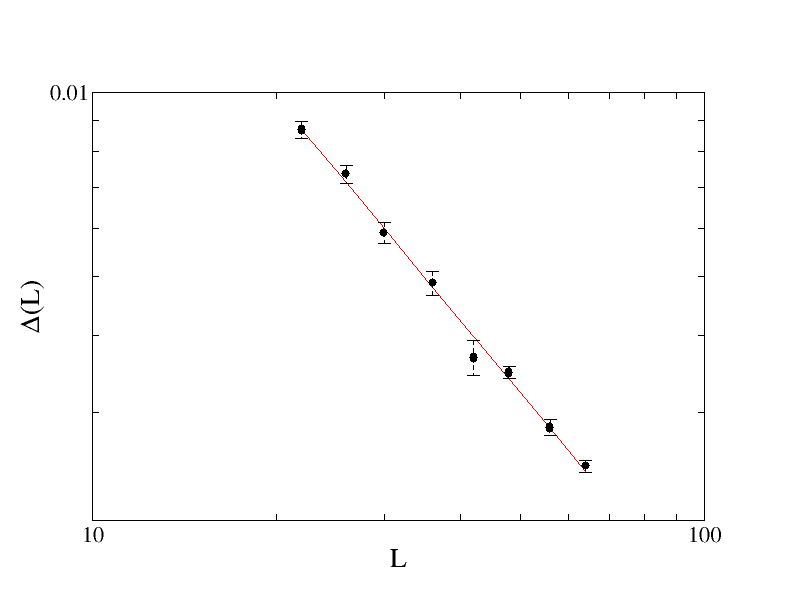}
\caption{Width of the probability to percolate $\Delta(L)$ as a function of system size $L$. The line is the best power-law fit, with slope $\frac{1}{\nu}=-1.205\pm0.038$.}
  \label{fgr:DL_L}
\end{figure}

\begin{figure}[h]
\centering
  \includegraphics[height=6.5cm]{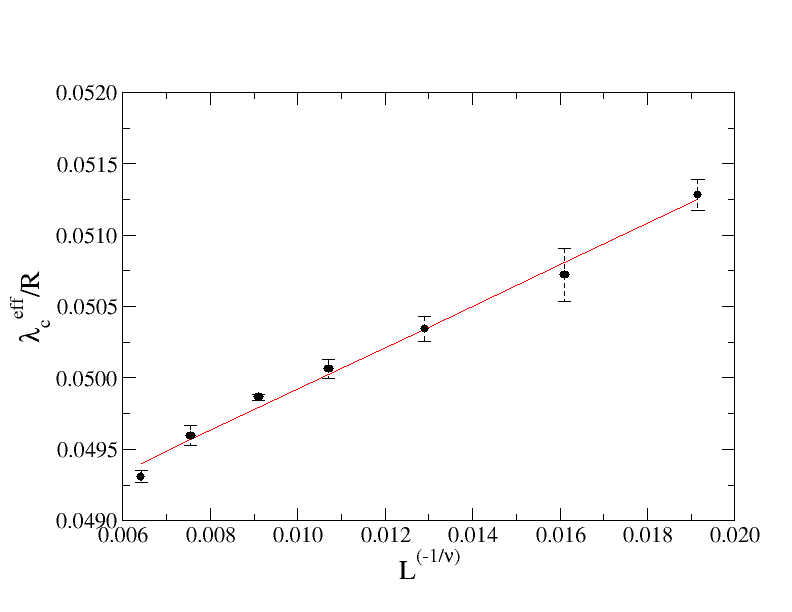}
\caption{Effective critical half-length of a capillary bridge length $\lambda_c^{eff}(L)$ against $L^{-\frac{1}{\nu}}$. The linear fit (continuous line) estimates a cut with the vertical axis at $\lambda_c=(0.049\pm 0.004)R$. }
  \label{fgr:Pc}
\end{figure}

The critical exponents and the critical half-length $\lambda_c$ for the transition are obtained by the finite-size scaling procedure proposed by Rintoul and Torquato \cite{Rintoul_Prec1997}. 
The scaling procedure starts by fitting the function 
\begin{equation} \label{eq:flD}
   	f(\lambda_c^{eff},\Delta(L))=[1 + \tanh[(\lambda - \lambda_c^{eff}(L))/\Delta(L)]]/2
\end{equation}
to the cumulative distribution of each size. The effective critical half-length $\lambda_c^{eff}(L)$ and the width $\Delta(L))$ of the distribution obtained from those fittings are plotted against the system size $L$ to obtain the critical parameters of the transition. Assuming that
\begin{equation} \label{eq:DL}
   	\Delta(L) \propto L^{\frac{1}{\nu}}\quad ,
\end{equation} 
and plotting $\Delta(L)$ against $L$ (Fig \ref{fgr:DL_L}) gives the critical exponent $\nu$ driving the divergence of the correlation length, $\xi\propto \left[\lambda-\lambda_c \right]^{-\nu}$. We obtain $\nu=0.830\pm 0.051$. Similarly, plotting $\lambda_c^{eff}(L)$ against $L^{-1/\nu}$ (Fig \ref{fgr:Pc}) estimates the critical value $\lambda_c$ for a system of infinite size, because
\begin{equation} \label{eq:betta}
   	\lambda_c^{\rm eff}(L)-\lambda_c \propto L^{-\frac{1}{\nu}}\quad .
\end{equation} 
Our result is  $\lambda_c=(0.049\pm 0.004)R$.

\section{Conclusions and Discussions}
\label{SecConclusions}

This work investigates the capillary ascent of a liquid through a granular soil as a percolation transition driven by the half maximal distance between two grains that can be overpassed by a capillary bridge, $\lambda$. Below a critical value $\lambda_c$ the structure of trimers and enclosed volumes is disconnected, and no water can rise. Above, the water percolates trough the sample. By using standard techniques of percolation analysis, we found $\lambda_c=(0.049\pm 0.004)R$ and a critical exponent for the correlation length $\nu=0.830\pm 0.051$. These results deserve some discussion. First, the finite-size scaling is of very good quality (Fig. \ref{fgr:DL_L}), suggesting that the technique fits the problem and that there is, indeed, a phase transition. Second, the critical length of a bridge, $2\lambda_c=0.098(6)R$, is less than one half of the mean distance between neighboring grains in the samples (around $(0.258 \pm 0.007)R$) for a volume fraction $\phi\simeq 0.63$), which can be considered as a naive {\it a priori} estimation of such parameter. Third, the critical exponent $\nu$ is, within error bars, the one of ordinary percolation ($\nu=0.87619(12)$ \cite{Xu_Simu2014}). This result suggests that each facet among three neighboring grains could turn into a trimer at disposal for the capillary rising almost independently from the other facets, a consequence of the random positions for the grains. This similarity should be confirmed by computing other critical exponents, like the critical exponent $\beta$ for the order parameter, which would be the probability of a cell to belong to the percolating cluster, as usual. Similarly, our study could be reproduced for other void fractions to determine how general are our conclusions. All these are thema for future works.  

An usual way to control capillary rising, as we mentioned before, consists in treating some part of the grains with hydrophobic solutions that alter the contact angle, reducing $\lambda$ (Fig. \ref{fgr:TorAp}). The key question here is which proportion of the material has to be treated to avoid capillary rising. The simulation would proceed in a similar way as the present one, but with three maximal lenghts $\lambda$: one between untreated grains, other between treated grains and a third one between one treated and one untreated grain. Such a future work will be of great interest in geotechnique.

The present study combines capillary structures and percolation theory to investigate capillary rising through a granular medium. It constitutes a novel approach and a new step in the understanding of this rich phenomenon.

\section{Acknowledgments}
\label{SecAcknowledgments}

We thanks COLCIENCIAS Young Researchers Program, Grant 2014-645, the Universidad Nacional de Colombia and the Portuguese Foundation for Sciences and Technology (FCT) under Contracts nos. UID/FIS/00618/2013, EXCL/FIS-NAN/0083/2012, and IF/00255/2013 for finnancial support.

%
%
%
\bibliography{Manuel_Cardenas_BIB} 

\begin{thebibliography}{25}

\bibitem{_Unsa2014}
N.~Khalili, A.~Russell, E.~Khoshghalb, A, \emph{Unsaturated soils : research
  and applications}, 1st~edn. (Crc Press, 2014)

\bibitem{Pozzato_Char2014}
A.~Pozzato, A.~Tarantino, Unsaturated Soils: Research and Applications
  \textbf{2}, 1082 (2014)

\bibitem{Fatt_The_1956}
I.~Fatt, Trans AIME \textbf{207}, 144 (1956)

\bibitem{Vogel_Quant_2001}
H.~Vogel, K.~Roth, Advances in Water Resources \textbf{24}, 233 (2001)

\bibitem{Coles_Pore1998}
M.~Coles, P.~Hazlett, W.~Soll, E.~Muegge, K.~Jones, Journal of Petroleum
  Science and Engineering \textbf{19}, 55 (1998)

\bibitem{Fischer_Pred_1999}
U.~Fischer, M.~Celia, Water Resour. Res. \textbf{6}, 1089 (1999)

\bibitem{Mason_Simu1995}
G.~Mason, D.~Mellor, Journal of Colloid and Interface Science \textbf{176}, 214
  (1995)

\bibitem{Scheel_Morp2008}
M.~Scheel, R.~Seemann, M.~Brinkmann, M.~Di~Michiel, A.~Sheppard,
  B.~Breidenbach, S.~Herminghaus, Nature Materials \textbf{7}, 189 (2008)

\bibitem{Motealleh_Unif2013}
S.~Motealleh, M.~Ashouripashaki, D.~DiCarlo, S.~Bryant, Transport in Porous
  Media \textbf{99}, 581 (2013)

\bibitem{Melnikov_Grai2015}
K.~Melnikov, R.~Mani, F.~Wittel, M.~Thielmann, H.~Herrmann, Cond-mat. Soft
  \textbf{92} (2015)

\bibitem{Megias_Capil2009}
D.~Megias-Alguacil, L.~Gauckler, AlChE \textbf{55} (2009)

\bibitem{Martys_Crit1991}
N.~Martys, M.~Robbins, Phys. Rev. Lett. \textbf{66}, 1058 (1991)

\bibitem{Cieplak_Dyna1988}
M.~Cieplak, O.~Robbins, Phys. Rev. Lett. \textbf{60}, 2042 (1988)

\bibitem{Berkowitz_Perc1993}
B.~Berkowitz, I.~Balberg, Water Resources Research \textbf{29}, 775 (1993)

\bibitem{Stauffer1994}
D.~Staufer, A.~Abarony, \emph{{I}ntroduction to percolation theory}, 2nd~edn.
  (Taylor and Francis, 1994)

\bibitem{Sahimi_Appl1994}
M.~Sahimi, \emph{{A}pplications of {P}ercolation {T}heory}, 1st~edn. (Taylor \&
  Francis, UK, 1994)

\bibitem{Araujo_Rece2014}
N.~Araujo, P.~Grassberger, B.~Kahng, K.~Schrenk, R.~Ziff, The European Physical
  Journal Special Topics \textbf{223}, 2307 (2014)

\bibitem{Guopingcap4}
G.~Lian, \emph{{C}omputer simulation of moist agglomerate colisions} (Doctoral
  Thesis, The University of Aston in Birmingham, 1994)

\bibitem{Fisher1926}
R.~Fisher, J. of Agricultural Science \textbf{16}, 492 (1926)

\bibitem{Shahinpoor_An1980}
M.~Shahinpoor, Powder Technology \textbf{25}, 163 (1980)

\bibitem{Jodrey_Comp1985}
J.~Jodrey, E.~Tory, Physical Review A \textbf{32} (1985)

\bibitem{Hawick_Para2010}
K.~Hawick, A.~Leist, D.~Playne, Parallel Computing \textbf{36}, 655 (2010)

\bibitem{Press_Nume2007}
W.~Press, S.~Teukoisky, W.~Vetterlingand, B.~Flannery, \emph{{N}umerical
  recipes}, 3rd~edn. (Cambridge, 2007)

\bibitem{Rintoul_Prec1997}
M.~Rintoul, S.~Torquato, J. Phys. A: Math. Gen. \textbf{30}, 585 (1997)

\bibitem{Xu_Simu2014}
X.~Xu, J.~Wang, J.~Lv, Y.~Deng, Frontiers of Physics \textbf{9}, 113 (2014)

\end{thebibliography}
%
%

\end{document}